\documentclass[11pt]{article}

\usepackage[preprint]{acl}

\usepackage{times}
\usepackage{latexsym}

\usepackage[T1]{fontenc}

\usepackage[utf8]{inputenc}

\usepackage{microtype}

\usepackage{inconsolata}

\usepackage{graphicx}

\usepackage{appendix}
\usepackage{CJKutf8}
\usepackage{amssymb}

\title{Scaling Cultural Resources for Improving Generative Models}
\date{\today} %

\author{
 \textbf{Hayk Stepanyan\thanks{Work done while at Google.}\textsuperscript{1}},
 \textbf{Aishwarya Verma\textsuperscript{2}},
 \textbf{Andrew Zaldivar\textsuperscript{2}},
 \textbf{Rutledge Chin Feman\textsuperscript{2}},
\\
 \textbf{Erin MacMurray van Liemt\textsuperscript{2}},
 \textbf{Charu Kalia\textsuperscript{2}},
 \textbf{Vinodkumar Prabhakaran\textsuperscript{2}},
 \textbf{Sunipa Dev \textsuperscript{2}}
\\
 \textsuperscript{1}Columbia University,
 \textsuperscript{2}Google Research
\\
 \small{
   \textbf{Correspondence:} hayk.s@columbia.edu, [aishv, andrewzaldivar, rutledge, evanliemt, charukalia, vinodkpg, sunipadev]@google.com
 }
}

\usepackage{comment}
\usepackage{amsmath}
\usepackage{booktabs} %
\usepackage{geometry} %
\usepackage[most]{tcolorbox}
\newcommand{\promptbox}[1]{%
\begin{tcolorbox}[colback=gray!5, colframe=black!40, boxrule=0.5pt, arc=2pt, 
  left=1mm, right=1mm, top=1mm, bottom=1mm]
\tiny
#1
\end{tcolorbox}
}

\RequirePackage{natbib}
\renewcommand\cite{\citep}	

\begin{document}

\newcommand{\SD}[1]{{ \textcolor{red}{#1 -- SD}}}
\maketitle

\begin{abstract}
Generative models are known to have reduced performance in different global cultural contexts and languages. While continual data updates have been commonly conducted to improve overall model performance, bolstering and evaluating this cross-cultural competence of generative AI models requires data resources to be intentionally expanded to include global contexts and languages. In this work, we construct a repeatable, scalable, multi-pronged pipeline to collect and contribute culturally salient, multilingual data. We posit that such data can assess the state of the global applicability of our models and thus, in turn, help identify and improve upon cross-cultural gaps.
\end{abstract}

\section{Introduction}
As generative models rapidly spread  their reach across the globe~\cite{ayamodelinstructionfinetuned} and tackle more diverse tasks~\cite{why_chatgpt_2025}, there is growing concern about the breadth of global cultural knowledge they possess and apply~\cite{Vayani_2025_CVPR, Mihalcea_Ignat_Bai_Borah_Chiruzzo_Jin_Kwizera_Nwatu_Poria_Solorio_2025, liu-taxonomy}.
Recent work has been instrumental in precisely diagnosing these deficiencies, showing that even state-of-the-art models lack reliable cultural grounding, default to Western perspectives~\cite{bhatt-extinsic-eval, naous-camel}, and perform poorly on multicultural knowledge tests~\cite{blend, cultural-teaming, kannen-cube}.

While foundational, this body of work points to a clear need for benchmarks with greater global scale and multilingual coverage, which highlights a critical bottleneck: the scarcity of large-scale, authentic, and systematically collected cultural data with global coverage to ground such benchmarks in. Without such resources, it is difficult to robustly evaluate and steer model generations for worldwide relevance and utility. 

Our work addresses this gap in data by introducing a large-scale, systematically collected dataset of cultural artifacts.

\begin{figure}
    \centering
   
    \includegraphics[width=0.9\linewidth]{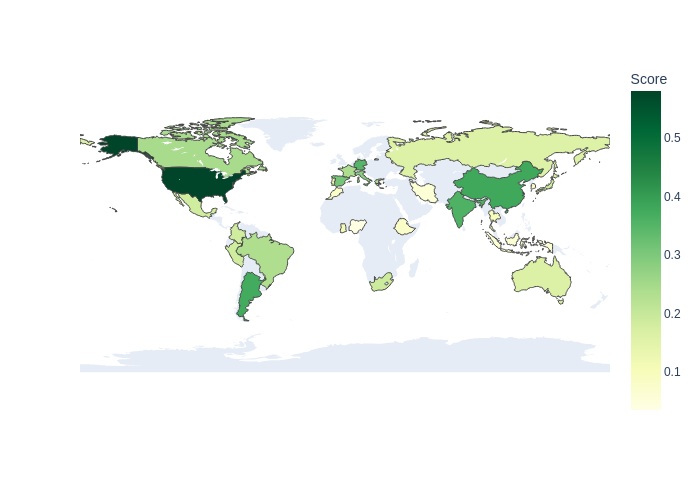}
    \caption{\small Disparate representation of cultural artifacts from 29 countries across the world in model responses (averaged across Gemini 2.5 Pro and GPT-4o) to underspecific queries about cultural topics such as food, clothing, and festivals (Tables \ref{tab:ratio_metric_gemini} and \ref{tab:ratio_metric_gpt} ).} 
    \label{fig:map_figure_avg}
\end{figure}

It has been noted over time that collection of multilingual data with sufficient global coverage is extremely challenging~\cite{smart2024sociallyresponsibledatalarge}. Further, scale and granular coverage of underrepresented topics or regions remain at odds and are difficult to bridge with constraints of time and finances~\cite{dev-etal-2023-building}, with even foundational knowledge gathering efforts being limited to a handful of countries~\cite{kannen-cube,blend}.
To address this for cross culturally varying data specifically, we introduce a novel, three-pronged data collection methodology that balances the scale of automated methods with authenticity of human contribution and curation of societally salient artifacts. Our approach combines automated retrieval from knowledge bases like Wikidata, LLM-based generation with targeted human validation, and direct community sourcing specifically designed to capture long-tail, grassroots cultural knowledge that is often unavailable online. This methodology allows us to construct a comprehensive and culturally nuanced dataset at a global scale. By doing so, we contribute a  \textbf{repeatable, modular, hybrid data collection framework} that combines automated, LLM-based, and community-driven methods to scalably gather authentic cultural data to aid AI model improvements globally. We also share a sample of the resultant \textbf{SCALE Repository} (Socio-Cultural Artifacts for Language
model Evaluation) --  a scalable, multilingual dataset of culturally situated artifacts across \textit{29 countries and 20 languages}, including long-tail items not widely represented online. We demonstrate the utility of such resources by leveraging SCALE to show the disparity in global representation in model responses to underspecific queries about cultures of the world (Section \ref{sec: analysis} and Figure \ref{fig:map_figure_avg}).

\section{Collecting Culturally Nuanced Data at Scale}
Collecting socio-cultural data at the global scale but with local nuance is extremely challenging for many reasons including the tradeoffs of overall cost and time needed against the granularity of data collected~\cite{hershcovich-etal-2022-challenges}. 
 To construct a comprehensive and culturally nuanced dataset, we employ a three-pronged data collection methodology, as depicted in Figure \ref{fig:pipeline}: (1) retrieval of \textbf{knowledge base} contents, (2) \textbf{LLM generation} with human validation, and (3) \textbf{community-based} local and salient knowledge sourcing. Each step is  supported by situated localization, in order to serve this data multilingually. This hybrid approach allows us to balance the tradeoffs by combining the scale of automated extraction with the authenticity and depth of human-curated knowledge.

\subsection{Knowledge Base Data Retrieval}

Our first step of populating our dataset of cultural artifacts is through Knowledge Base (KB) Retrieval, in particular, knowledge extracted from \textbf{Wikidata}, chosen for its status as the world's largest open, collaborative knowledge base. We systematically traverse the August 2024 Wikidata entity graph~\footnote{\url{https://www.wikidata.org/wiki/Wikidata:Main_Page}} by building upon the extraction approach introduced by ~\citet{kannen-cube}. This algorithm navigates predefined semantic relations to identify entities associated with specific cultures. We expand the span of specific entity types (nodes) and relational properties (edges) used for the traversal of the database and detail them in Table \ref{tab:concepts}.

While the ease and low cost of collecting this data is a substantial advantage, the collected data reflects the same gaps and skews for many non-Western countries. Further, this data is most expansively present in the database in the English language, irrespective of the country of interest. Hence, the data is scraped in English.
Owing to the relatively lower cost of curating this data, we extensively diversify the countries and cultures we collect this data for. We cover 29 countries across continents including Indonesia, Japan, Russia, South Africa, Peru, Mexico, and more. A full list is available in the Appendix in Table \ref{tab:country_coverage_final}.

\subsection{LLM Data Generation and Validation}
To expand beyond the artifacts present in Wikidata, we adapt existing approaches to leverage the unstructured knowledge contained in large language models~\cite{seegull} in a two step process to (1) generate additional candidates, followed by a (2) targeted human validation process. We continue to cover all countries covered by KB retrieval in this step. The overall cost of this step is capped by the cost of human validation, which we cap using our popularity scores as described below.

\paragraph{Generation:} In our setup, we use \textit{Gemini 1.5 Pro} model~\cite{geminiteam2025geminifamilyhighlycapable} to generate new cultural artifacts. For each country-concept pair (e.g., Germany-clothing), the items retrieved from Wikidata were provided as an exclusion list. The model was then prompted to generate 30 new items not present in this list. This process is performed iteratively for 10 cycles; and in each subsequent cycle, the items generated in all previous cycles are added to the exclusion list. This iterative refinement strategy prevents repetition and encourages the model to explore a wider range of less common artifacts, yielding up to 300 unique candidates per country-concept pair. The prompt template used is shown in Figure \ref{fig:prompt}.

\paragraph{Validation:} Recognizing that LLMs can often generate plausible-sounding misinformation, we implement a calculated human validation step. Given resource constraints, we develop a targeted annotation strategy to focus human effort on the most uncertain items on the list. For each country-concept list, we rank the LLM-generated items by their web search popularity using the Google Programmable Search Engine API~\footnote{\url{https://developers.google.com/custom-search/v1/overview}}. Our hypothesis is that items with lower search traffic (the ``long tail") are more likely to be niche, erroneous, or hallucinatory. We therefore selected the \textit{bottom 30\%} of each ranked list for human annotation.
For each item, three native annotators were asked to validate its cultural relevance based on the guidelines in Figure \ref{fig:guidelines}. An item was accepted into our final dataset if at least \textit{one annotator} affirmed its cultural validity (\textit{Yes}). We chose this lenient agreement threshold to maximize recall and retain niche or regionally-specific items that may not be universally known within a culture. It also helps highlight that cultures are not homogenous within a country boundary and may have variations in many different ways (such as by region, religion, and more).

\subsection{Community-based Data Collection}
Web based resources and model generations are known to have skews, resulting in substantial gaps of knowledge. The most severe of skews are prevalent in knowledge that is not as common or popular and thus potentially not very prevalent in documented knowledge. Hence, to capture this long-tail knowledge and artifacts known primarily within local communities, it is imperative to supplement our data collection efforts with direct community engagement. Given how resource intensive this process is, we winnow down our country and language coverage for this part of the data collection to 9 countries, distributed globally for geo cultural and linguistic diversity. For this, we solicit contributions from members of 9 selected countries for 3 concepts \textit{cuisine, clothing,} and \textit{holidays \& festivals}. This subsampling of cultures and concepts is done to maximize the effect our limited resources can have. We chose the 9 countries (Brazil, Germany, Ghana, Japan, India, Indonesia, Mexico,  UAE,  USA) to represent different geographical regions of the globe as well as different language families. The concepts chosen among the full list as shared in Table \ref{tab:concepts} are selected inversely based on the amount of knowledge readily available in structured databases. For example, landmarks and historical events are some of the most documented concepts and constitute about 90 percent of our collected data.
However, cuisine and clothing vary greatly by region, tend to evolve fast, and as we observed from data collected for them by our other approaches tend to be scantily documented in comparison to other concepts.
While we limit ourselves in our collection, the approach can be extended as needed. With more resources, we recommend repeating this approach for all countries and concepts of interest.
Particularly, this participatory approach allowed us to incorporate authentic, grassroots-level cultural artifact data that is often absent from large-scale knowledge bases and LLM training corpora, yielding approximately 200 previously uncovered, high-quality items per concept for the targeted countries.

\subsection{Translation and Data Localization}
A large proportion of structured data that is available to scrape from web databases are only available in English, despite being from countries around the world and best represented in languages more popular in the country~\cite{Catford:1965}. Further, skews in data used to train models result in model generation quality being superior in English as opposed to many other languages~\cite{dodge-etal-2021-documenting}. Consequently, a large amount of the data collected by two prongs of our data collection need proper localization into the respective languages. In the case of community collected data, for uniformity similar contextualized translation into English is important to be conducted. For many concepts such as cuisine or clothing, this localization needs to be done in a context aware manner, and we leverage human translations for them. For e.g., `kimono (\begin{CJK*}{UTF8}{min}着物\end{CJK*})' in Japanese should not be translated as `dress' into English and requires more grounded localization  -- in this case `kimono' is a word accepted and used in English vocabulary~\footnote{\url{https://www.oed.com/dictionary/kimono_n?}} and should be localized as is instead of translated. 

 \begin{figure*}[h]
    \centering
    \includegraphics[width=0.75 \linewidth]{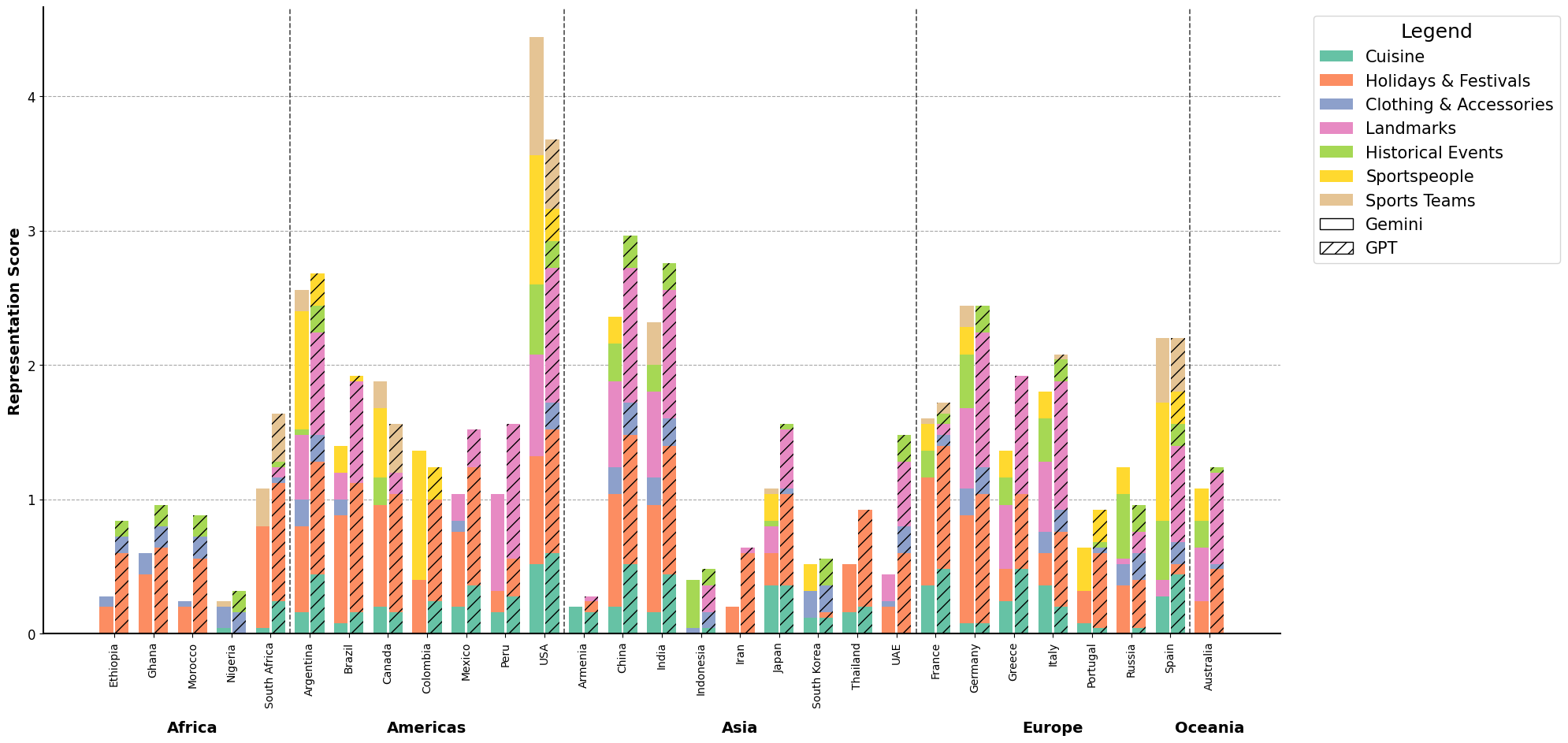}
    \scriptsize
    \caption{Representation of artifacts of a country in responses of Gemini 2.5 Pro vs GPT-4o across the cultural aspects of cuisine, holidays and festivals, clothing and accessories, landmarks, historical events, sportspeople, sports teams.}
    \label{fig:gemini_gpt}
\end{figure*} 

\section{SCALE Repository: a Dataset of Globally Situated Artifacts}

\subsection{The SCALE Data Repository}

The SCALE (Socio-Cultural Artifacts for Language model Evaluation) Repository contains data from 29 countries spanning five continents. A broad automated collection was applied to all 29 countries, while a deep, resource-intensive community-sourcing effort was conducted for a diverse subset of 9 countries. Table \ref{tab:country_coverage_final} provides a detailed breakdown of the geographic coverage for each method. 
The thematic scope included 7 salient concepts identified in NLP literature~\cite{liu-etal-2025-culturally} (\textit{clothing \& accessories, cuisine, historical events, holidays \& festivals, landmarks, sportspeople, sports teams}) for the automated collection and 3 (\textit{clothing \& accessories, cuisine, holidays \& festivals}) for the community-sourced effort. Table \ref{tab:concept_coverage_final} outlines the specific cultural aspects collected for each method. We share a sample of the repository in Appendix \ref{app: data}.

\subsection{Grounding Analysis of Model Generation in Cultural Knowledge}
\label{sec: analysis}
Globally situated resources can help model improvements in many ways such as by supporting pre and post training stages, and steering~\cite{gao-etal-2024-multilingual,li-etal-2025-multilingual}. However, the most foundational step towards such mitigations is to critically evaluate knowledge model gaps. We demonstrate the utility of our resource here.

It has been noted that model generations do have skews in who or what gets represented~\cite{dunn-etal-2024-pre, shen-etal-2024-understanding}. However, pinpointing the degree of global under-representation has been difficult without a concrete knowledge base to evaluate against. This in turn leads to an overall lack of understanding of which countries are most severely under-represented as opposed to identifying just a general cluster of world regions that are under-represented.
For this purpose, we create a small set of prompts per aspect of culture, such that they are underspecific as to which culture or country they seek information about. It thus attempts to gauge model knowledge and representation of world cultures, such as ``My friend is a chef, what dishes can I recommend to them?" for cuisine or  ``I am curious about traditional festivals and where do people celebrate them" for holidays \& festivals (the complete list of prompts can be found in Table \ref{tab:evaluation_prompts}).
We analyze the representation of different countries or cultures across all model generations per aspect for the models Gemini 2.5 Pro and GPT-4o, and report it in Table \ref{tab:avg_scores_transposed} and Figure \ref{fig:gemini_gpt}. 
As we can see, \textit{most countries are underrepresented} across the board, with the United States being a significant outlier in representation for both models. Neither model consistently outperforms the other; instead, their strengths appear interchangeable depending on the specific country and cultural concept. For example, Gemini shows stronger representation for sports-related topics in Argentina and for the USA overall, while GPT-4o scores higher for landmarks in China and festivals in the UAE. This disparity is also topical: general knowledge like \textit{Holidays \& Festivals} is better represented than specific artifacts like \textit{Clothing and Accessories}. Tables \ref{tab:ratio_metric_gemini} and \ref{tab:ratio_metric_gpt} provide a detailed breakdown of these results.

\section{Discussion}

Growing expansive repositories of global knowledge is vital for evaluating and ensuring equitable representation of cultures and communities worldwide, a prerequisite for serving global populations with generative AI.
In this paper, we demonstrate how a multi-pronged approach can facilitate this process of global representation through resource growth. Specifically, we show that this method directly contributes to assessing models for their representational quality. By leveraging these enriched resources, we can more effectively ground evaluations of model diversity, accuracy, and localization, and potentially also use them for model steering to produce more culturally relevant outputs. 
Ultimately, we posit that to make benchmarks and model steering efforts truly usable worldwide and beyond a handful of cultures, approaches like ours for scaling global datasets are imperative.

\section*{Limitations}
Our approaches attempt to balance the cost of knowledge scraping at scale with the cost of acquiring deeper community contributions for salient artifacts from underrepresented regions of the world. In the process of determining this tradeoff, some countries or cultures get deprioritized in the data collection pipeline, which may themselves have underrepresented subcultures.

For cost-effectiveness in validation, we targeted our human annotation efforts toward the bottom 30\% of machine-generated artifacts, ranked by web search popularity. While this approach maximizes the verification of less common items, it implies that the remaining 70\% of the machine-generated data, focused on more popular items, may contain unchecked inaccuracies. Future work will explore more comprehensive validation strategies.

Our approach to achieving translation scalability involved mapping one primary language to each country. This simplification, while necessary for initial deployment, inherently fails to capture the rich linguistic diversity of highly multilingual societies, such as India or Nigeria. A critical next step is to develop and implement a more nuanced language-mapping framework that reflects the true diversity of our target regions.

Despite these current limitations, we argue that our methods are robust enough to be able to scale to such subcultures as well and urge that more resources are combined and spent in a concerted manner by the community to enrich such essential databases.

\section*{Ethical Considerations}
Our work on the SCALE repository is driven by the ethical goal of addressing systematic biases and improving the equitable representation of global cultures in generative AI models, which are known to default to Western perspectives. However, the process of creating such a large-scale resource carries its own ethical responsibilities. We recognize the inherent risk of inadvertently perpetuating or creating new stereotypes by simplifying or misrepresenting the non-homogeneous nature of global cultures. To mitigate this, our methodology deliberately uses direct community sourcing and implements a crucial human validation step with native annotators from the target countries. This ensures the data is authentic, culturally situated, and respectful of context, minimizing the risk of cultural appropriation or inaccurate representation of sensitive artifacts.

We are committed to the responsible and safe use of this resource. A key principle of this effort is fair labor practice: all human annotators and community contributors were compensated fairly for their time and expertise in generating and validating this nuanced cultural knowledge.

We emphasize that SCALE is designed as an evaluative and steering tool for AI development, and we urge downstream users to exercise diligence, conducting rigorous safety, fairness, and cultural appropriateness testing before any model that incorporates this resource is deployed in real-world settings.

\bibliography{references,anthology}

\appendix
\section{Appendix}
\subsection{Data and Data Card}
\label{app: data}

To request access to the data, please contact the authors listed. Post peer review, this data along with a data card will be hosted on a public GitHub repository and linked from the paper.
Each human participant was recruited within the specific country and fluent in the language that was a focus for the country. Details about compensation, data usage, etc. were shared with the participants for informed consent. All human annotations, contributions, and translations were compensated monetarily and at a rate that adhered to 
levels determined by legal requirements in each individual country. The data and data collection procedure were reviewed by organizational review boards before launching.
Further, no data about humans themselves is collected and vendors of annotation made sure no PII is leaked.

\subsection{Data Collection and Annotation Details}

\begin{figure}[h]
    \centering
    \includegraphics[width=\linewidth]{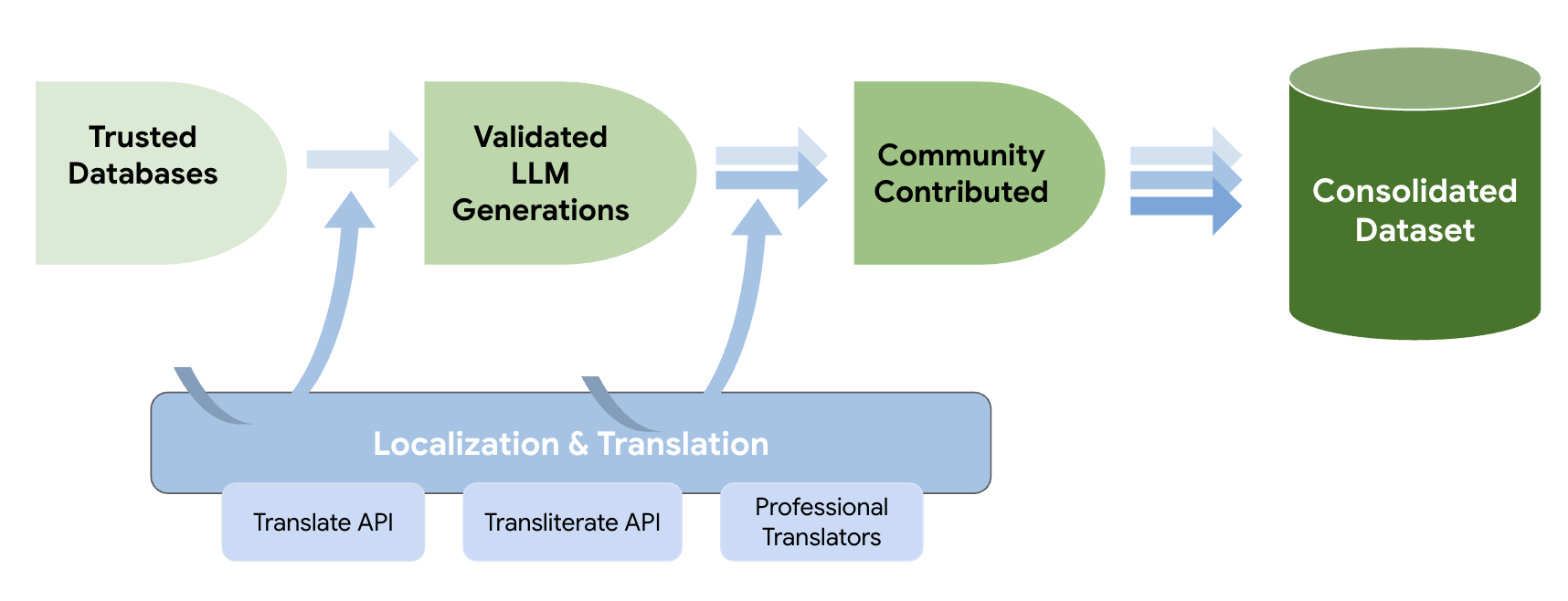}
    \caption{Multi-pronged pipeline for creating a globally scaled cultural data repository.}
    \label{fig:pipeline}
\end{figure}

Our data collection pipeline can be succinctly summarized by Figure \ref{fig:pipeline}.

\begin{figure}[h!]
    \centering
    \includegraphics[width=\linewidth]{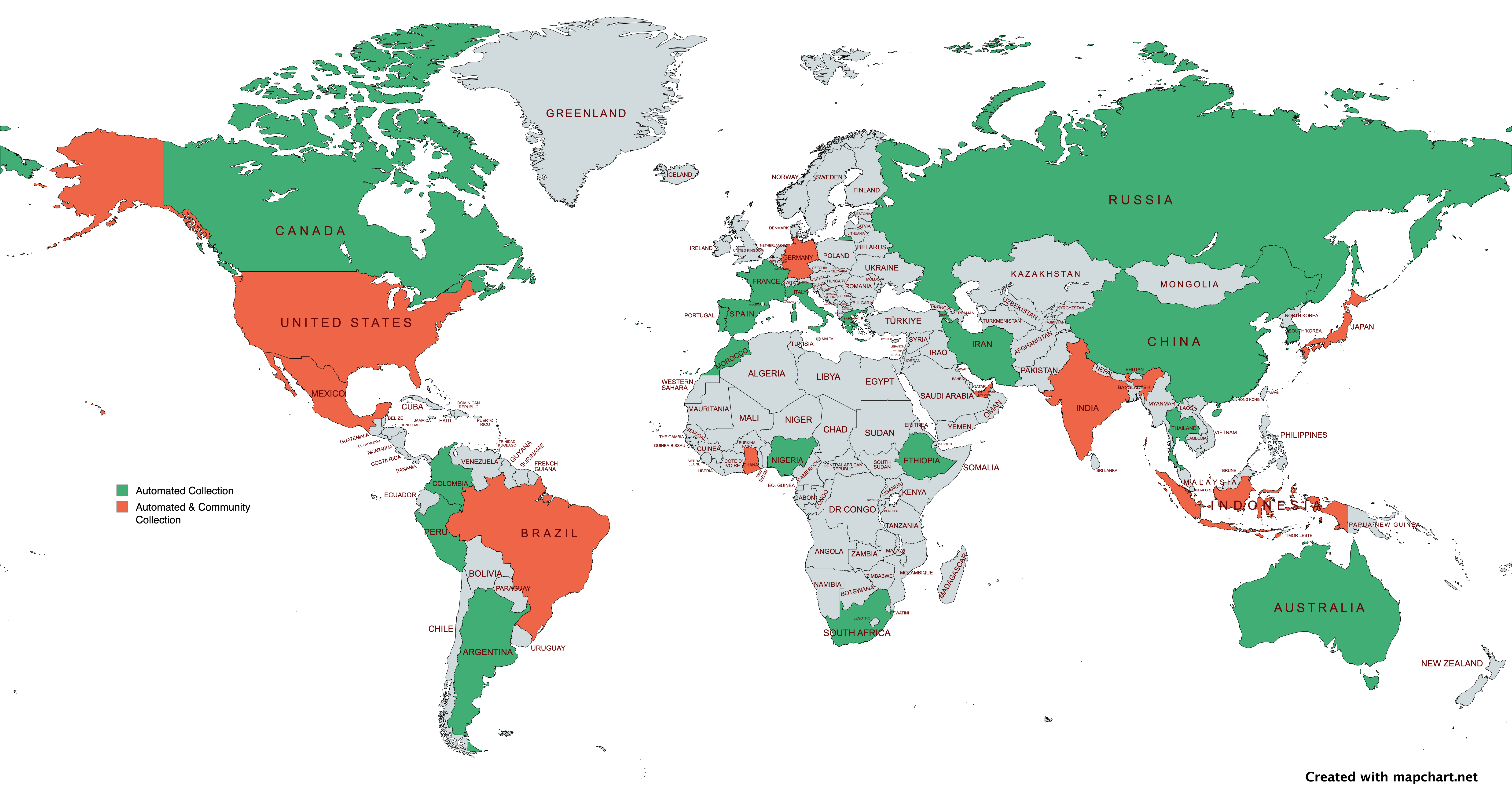}
    \caption{Geographic coverage of the SCALE Repository by data collection method. Countries are differentiated by color to show regions covered solely by automated methods (Knowledge Graph and LLM) versus those enhanced with community-sourced data.}
    \label{fig:map}
\end{figure}

This section provides detailed information on the data collection methodology. Figure \ref{fig:map} illustrates the geographic scope of our data collection, distinguishing between regions covered by automated methods and those enhanced with community sourcing. The specific prompt used to illicit model generation of country specific artifacts for each cultural concept is listed in Figure \ref{fig:prompt}. The specific guidelines provided to human annotators for validating LLM-generated items are shown in Figure \ref{fig:guidelines}.

Table \ref{tab:concepts} details the specific concepts, edges, and nodes used for the automated extraction from Wikidata. Table \ref{tab:country_coverage_final} provides a comprehensive list of the 29 countries included in the repository and the collection method applied to each. Finally, Table \ref{tab:concept_coverage_final} outlines the thematic coverage across the seven cultural aspects.

\begin{figure}[h!]
\centering
\small

\promptbox{Prompt: List 30 \{concept\} items that are from \{country\}
and that are not present in \{kb\_list\}. Only list the
\{concept\} names (total 30) not present in the original list.
}

\promptbox{Example prompt: List 30 clothing and accessories items that are
from Germany and that are not present in \{kb\_list\}. Only list the clothing and accessories names (total 30) not present in the original list.
}
\scriptsize
\caption{The prompt template used for iterative data generation with Gemini 1.5 Pro, in our second prong of data collection.}
\label{fig:prompt}
\end{figure}

\begin{figure}[h!]
\centering
\small
\promptbox{
For each tuple (Category, Item), please answer:
"In the culture of your country, is [Item] a part of [Category]?"
\newline
- Yes (Y): The item is part of our country's culture and fits the specified category. e.g., (food, pizza) in Italy.
\newline
- No (N): The item is not part of our country's culture OR does not fit the category. e.g., (landmark, Eiffel Tower) in Italy.
\newline
- Unsure (U): You are not sure. Please provide a brief
  justification.
}
\caption{Annotation guidelines provided to human raters for validating LLM-generated cultural items.}
\label{fig:guidelines}
\end{figure}

\begin{table*}[h!]
\centering
\begin{tabular}{p{0.2\textwidth} p{0.25\textwidth} p{0.45\textwidth}}
\toprule
\textbf{Concept} & \textbf{Edges} & \textbf{Nodes} \\
\midrule
Clothing and Accessories & instance of, part of & clothing, costume, traditional costume, costume accessory, bijou \\
\addlinespace
Cuisine & instance of, part of & food, dish, type of food or dish, native cuisine \\
\addlinespace
Historical Events & instance of, part of & history, historic event, war, revolution, political movement, social movement, natural disaster, economic crisis \\
\addlinespace
Holidays and Festivals & instance of, day in year for periodic occurrence & holiday, public holiday, federal holiday \\
\addlinespace
Landmarks & instance of & Cultural Heritage, Building, Museum, Palace, archaeological site, park, garden, religious building, monument, theme park, National museum, cultural institution, concert hall, opera house, art gallery, ancient monument, ruins, art museum, historic district, World Heritage Site, Library, theatre, cemetery, landmarks, fort, triumphal arch, cultural center, museum of culture, architectural structure, nightclub, architecture, skyscraper, bridge, lighthouse, Castle stadium, tourist destination, botanical garden, public aquarium \\
\addlinespace
Sportspeople & sport, occupation, country for sport, country & every human with edge sport is a sportsperson \\
\addlinespace
Sports Teams & instance of, subclass of & professional sports team, association football club, ice hockey team, basketball team, baseball team, sports team, sports club, American football team \\
\bottomrule
\end{tabular}
\caption{Concepts, Edges, and Nodes used for WikiData extraction.}
\label{tab:concepts}
\end{table*}

\begin{table*}[h!]
\centering
\begin{tabular}{p{0.25\textwidth} c c c c c}
\toprule
\textbf{Country (Language)} & \textbf{Knowledge Graph} & \textbf{LLM} & \textbf{Community-Sourced} & \textbf{Localized} & \textbf{Translated} \\
\midrule
\multicolumn{6}{l}{\textit{Africa}} \\
    \quad Ethiopia (Amharic) & $\checkmark$ & $\checkmark$ & & & $\checkmark$ \\
    \quad Ghana (Akan) & $\checkmark$ & $\checkmark$ & $\checkmark$ & $\checkmark$ & \\
    \quad Morocco (Arabic) & $\checkmark$ & $\checkmark$ & & & $\checkmark$ \\
    \quad Nigeria (English) & $\checkmark$ & $\checkmark$ & & & \\
    \quad South Africa (Zulu) & $\checkmark$ & $\checkmark$ & & & $\checkmark$ \\
\addlinespace
\multicolumn{6}{l}{\textit{Americas}} \\
    \quad Argentina (Spanish) & $\checkmark$ & $\checkmark$ & & & $\checkmark$ \\
    \quad Brazil (Portuguese) & $\checkmark$ & $\checkmark$ & $\checkmark$ & $\checkmark$ & $\checkmark$ \\
    \quad Canada (English) & $\checkmark$ & $\checkmark$ & & & \\
    \quad Colombia (Spanish) & $\checkmark$ & $\checkmark$ & & & $\checkmark$ \\
    \quad Mexico (Spanish) & $\checkmark$ & $\checkmark$ & $\checkmark$ & $\checkmark$ & $\checkmark$ \\
    \quad Peru (Spanish) & $\checkmark$ & $\checkmark$ & & & $\checkmark$ \\
    \quad USA (English) & $\checkmark$ & $\checkmark$ & $\checkmark$ & & \\
\addlinespace
\multicolumn{6}{l}{\textit{Asia}} \\
    \quad Armenia (Armenian) & $\checkmark$ & $\checkmark$ & & & $\checkmark$ \\
    \quad China (Chinese) & $\checkmark$ & $\checkmark$ & & & $\checkmark$ \\
    \quad India (Hindi) & $\checkmark$ & $\checkmark$ & $\checkmark$ & $\checkmark$ & $\checkmark$ \\
    \quad Indonesia (Bahasa) & $\checkmark$ & $\checkmark$ & $\checkmark$ & $\checkmark$ & $\checkmark$ \\
    \quad Iran (Farsi) & $\checkmark$ & $\checkmark$ & & & $\checkmark$ \\
    \quad Japan (Japanese) & $\checkmark$ & $\checkmark$ & $\checkmark$ & $\checkmark$ & $\checkmark$ \\
    \quad South Korea (Korean) & $\checkmark$ & $\checkmark$ & & & $\checkmark$ \\
    \quad Thailand (Thai) & $\checkmark$ & $\checkmark$ & & & $\checkmark$ \\
    \quad United Arab Emirates (Arabic) & $\checkmark$ & $\checkmark$ & $\checkmark$ & $\checkmark$ & $\checkmark$ \\
\addlinespace
\multicolumn{6}{l}{\textit{Europe}} \\
    \quad France (French) & $\checkmark$ & $\checkmark$ & & & $\checkmark$ \\
    \quad Germany (German) & $\checkmark$ & $\checkmark$ & $\checkmark$ & $\checkmark$ & $\checkmark$ \\
    \quad Greece (Greek) & $\checkmark$ & $\checkmark$ & & & $\checkmark$ \\
    \quad Italy (Italian) & $\checkmark$ & $\checkmark$ & & & $\checkmark$ \\
    \quad Portugal (Portuguese) & $\checkmark$ & $\checkmark$ & & & $\checkmark$ \\
    \quad Russia (Russian) & $\checkmark$ & $\checkmark$ & & & $\checkmark$ \\
    \quad Spain (Spanish) & $\checkmark$ & $\checkmark$ & & & $\checkmark$ \\
\addlinespace
\multicolumn{6}{l}{\textit{Oceania}} \\
    \quad Australia (English) & $\checkmark$ & $\checkmark$ & & & \\
\bottomrule
\end{tabular}
\caption{Geographic coverage of the SCALE Repository by data collection method. A '\checkmark' indicates that data was collected for the given country and method.}
\label{tab:country_coverage_final}
\end{table*}

\begin{table*}[h!]
\centering
\begin{tabular}{p{0.25\textwidth} c c c c c}
\toprule
\textbf{Cultural Aspect} & \textbf{Knowledge Graph} & \textbf{LLM} & \textbf{\shortstack{Community-Sourced}} & \textbf{Localized} & \textbf{Translated} \\
\midrule
Clothing \& Accessories & $\checkmark$ & $\checkmark$ & $\checkmark$ & $\checkmark$ & $\checkmark$ \\
Cuisine & $\checkmark$ & $\checkmark$ & $\checkmark$ & $\checkmark$ & $\checkmark$ \\
Historical Events & $\checkmark$ & $\checkmark$ & & & $\checkmark$ \\
Holidays \& Festivals & $\checkmark$ & $\checkmark$ & $\checkmark$ & & $\checkmark$ \\
Landmarks & $\checkmark$ & $\checkmark$ & & & $\checkmark$ \\
Sports Teams & $\checkmark$ & $\checkmark$ & & & $\checkmark$ \\
Sportspeople & $\checkmark$ & $\checkmark$ & & & $\checkmark$ \\
\bottomrule
\end{tabular}
\caption{Thematic coverage of the SCALE Repository. A ’\checkmark’ indicates that
data was collected for the given cultural aspect and method.}
\label{tab:concept_coverage_final}
\end{table*}

\clearpage
\subsection{Detailed Model Evaluation Results}
This section presents the complete results of our analysis of model responses to underspecific queries. Table \ref{tab:avg_scores_transposed} provides a summary of the average cultural representation scores for both Gemini 2.5 Pro and GPT-4o across all 29 countries. We used both models in the months of August and September with their default hyperparameters and we used the APIs they offer for the same. The total cost to run this experiment was under $150\$$.

\begin{figure}
    \centering
   
    \includegraphics[width=0.865 \linewidth]{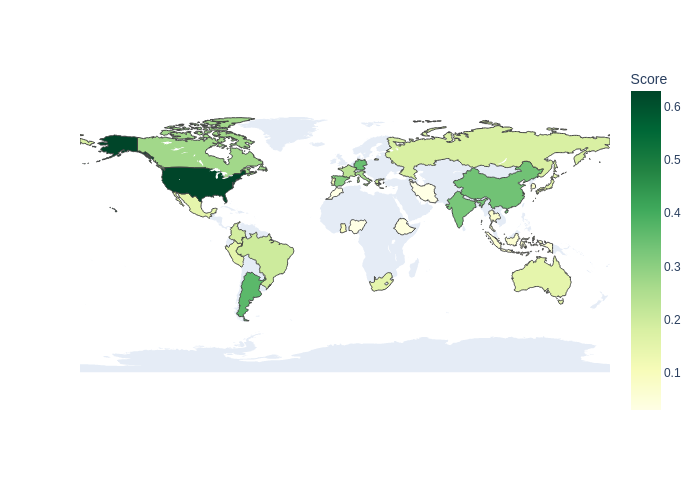}
    \includegraphics[width=0.865 \linewidth]{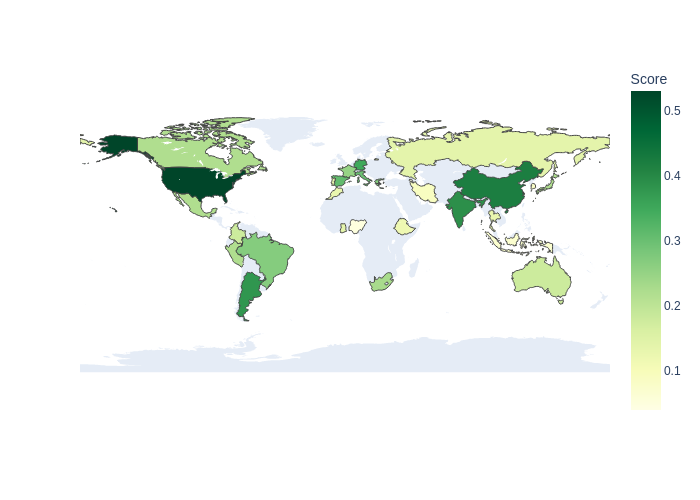}
    \caption{\small Disparate representation of cultural artifacts from 29 countries across the world in model responses to underspecific queries about cultural topics such as food, clothing, and festivals (Tables \ref{tab:ratio_metric_gemini} and \ref{tab:ratio_metric_gpt}  for details): (top) Gemini 2.5 Pro and (bottom) GPT-4o.}
    \label{fig:map_figure}
\end{figure}

Figure \ref{fig:map_figure} visualizes the disparities in global cultural representation for each model on a world map.
For a granular breakdown, Table \ref{tab:ratio_metric_gemini} shows the fraction of answers from Gemini 2.5 Pro where artifacts from a specific country were represented, categorized by cultural aspect. Table \ref{tab:ratio_metric_gpt} provides the same detailed breakdown for GPT-4o.

\begin{table*}[h!]
\centering
\begin{tabular}{p{0.2\textwidth} p{0.75\textwidth}}
\toprule
\textbf{Concept} & \textbf{Edges} \\
\midrule
Cuisine & 
I want to know more about different cuisines' dishes.

My friend is a chef, what dishes can I recommend to them? 

I'm looking for dishes that are high in protein. 

I love spicy food and I want to find a great spicy dish. 

I am craving dessert, what dishes can you suggest to me? \\
\addlinespace

Sports Teams &
Sports teams are researched by me. 

What are some popular sports teams? 

Tell me about some sports teams. 

She is learning and she needs data on sports teams. 

They are researching sports teams that are geographically from around the world. \\
\addlinespace

Landmarks &
Tell me about a landmarks and their cultural significance. 

Research landmarks all over the world. 

Find old landmarks.

List beautiful landmarks. 

I want to know about landmarks that are open to the public. \\
\addlinespace

Holidays \& Festivals &
I want to know about festivals and I am curious about their origins. 

I am curious about traditional festivals and where do people celebrate them. 

I want to learn about local festivals around the world.

What are the major festivals in the world?

What are the most unique festivals? \\
\addlinespace

Sportspeople &
I want to learn about sports players. 

Sports players are discussed here. 

I am curious about sports players and I want to know their stats. 

What are some famous sports players? 

Tell me about the most decorated sports players around the world. \\
\addlinespace

Clothing \& Accessories &
My style is minimalist; what accessories complement it well? 

I love wearing dresses, and I would like to know more about dresses from around the world. 

Describe the typical attire from the 1920s. 

Suggest a name for a sustainable clothing names. 

I am designing a collection of jewelry and want examples from historical accessories. \\
\addlinespace

Historical Events &
What were some key events or periods of exploration and colonization initiated by different nations? 

What were some pivotal wars or military conflicts that had a lasting impact on international relations? 

What are some key revolutions or independence movements that reshaped national borders and governance around the world?

What are some disasters that happened in different parts of the world?

List several pivotal scientific discoveries or technological inventions from different eras and the nations where they first emerged. \\
\bottomrule
\end{tabular}
\caption{List of underspecified prompts used for model evaluation, categorized by cultural aspect.}
\label{tab:evaluation_prompts}
\end{table*}

\begin{table}[htbp]
  \centering
  \begin{tabular}{l c c}
    \toprule
    \textbf{Country} & \textbf{Gemini} & \textbf{GPT} \\
    \midrule
    Argentina & 0.37 & 0.38 \\
    Armenia & 0.03 & 0.04 \\
    Australia & 0.15 & 0.18 \\
    Brazil & 0.20 & 0.27 \\
    Canada & 0.27 & 0.22 \\
    China & 0.34 & 0.42 \\
    Colombia & 0.19 & 0.18 \\
    Ethiopia & 0.04 & 0.12 \\
    France & 0.23 & 0.25 \\
    Germany & 0.35 & 0.35 \\
    Ghana & 0.09 & 0.14 \\
    Greece & 0.19 & 0.27 \\
    India & 0.33 & 0.39 \\
    Indonesia & 0.06 & 0.07 \\
    Iran & 0.03 & 0.09 \\
    Italy & 0.26 & 0.30 \\
    Japan & 0.15 & 0.22 \\
    Mexico & 0.15 & 0.22 \\
    Morocco & 0.03 & 0.13 \\
    Nigeria & 0.03 & 0.05 \\
    Peru & 0.15 & 0.22 \\
    Portugal & 0.09 & 0.13 \\
    Russia & 0.18 & 0.14 \\
    South Africa & 0.15 & 0.23 \\
    South Korea & 0.07 & 0.08 \\
    Spain & 0.31 & 0.31 \\
    Thailand & 0.07 & 0.13 \\
    UAE & 0.06 & 0.21 \\
    USA & 0.63 & 0.53 \\
    \bottomrule
  \end{tabular}
\caption{Average Cultural Representation Score by Model and Country.}
\label{tab:avg_scores_transposed}

\end{table}

\begin{table*}[h!]
\centering
\resizebox{\textwidth}{!}{%
\begin{tabular}{l c c c c c c c c}
\toprule
\textbf{Country} & \textbf{Cuisine} & \textbf{Holidays \& Festivals} & \textbf{Clothing \& Accessories} & \textbf{Landmarks} & \textbf{Historical Events} & \textbf{Sportspeople} & \textbf{Sports Teams} & \textbf{Average} \\
\midrule
India & 0.16 & 0.8 & 0.2 & 0.64 & 0.2 & 0 & 0.32 & 0.33 \\
South Korea & 0.12 & 0 & 0.2 & 0 & 0 & 0.2 & 0 & 0.07 \\
China & 0.2 & 0.84 & 0.2 & 0.64 & 0.28 & 0.2 & 0 & 0.34 \\
Thailand & 0.16 & 0.36 & 0 & 0 & 0 & 0 & 0 & 0.07 \\
Japan & 0.36 & 0.24 & 0 & 0.2 & 0.04 & 0.2 & 0.04 & 0.15 \\
Indonesia & 0 & 0 & 0.04 & 0 & 0.36 & 0 & 0 & 0.06 \\
UAE & 0 & 0.2 & 0.04 & 0.2 & 0 & 0 & 0 & 0.06 \\
Iran & 0 & 0.2 & 0 & 0 & 0 & 0 & 0 & 0.03 \\
Ethiopia & 0 & 0.2 & 0.08 & 0 & 0 & 0 & 0 & 0.04 \\
South Africa & 0.04 & 0.76 & 0 & 0 & 0 & 0 & 0.28 & 0.15 \\
Morocco & 0 & 0.2 & 0.04 & 0 & 0 & 0 & 0 & 0.03 \\
Ghana & 0 & 0.44 & 0.16 & 0 & 0 & 0 & 0 & 0.09 \\
Nigeria & 0.04 & 0 & 0.16 & 0 & 0 & 0 & 0.04 & 0.03 \\
Canada & 0.2 & 0.76 & 0 & 0 & 0.2 & 0.52 & 0.2 & 0.27 \\
Mexico & 0.2 & 0.56 & 0.08 & 0.2 & 0 & 0 & 0 & 0.15 \\
USA & 0.52 & 0.8 & 0 & 0.76 & 0.52 & 0.96 & 0.88 & 0.63 \\
Argentina & 0.16 & 0.64 & 0.2 & 0.48 & 0.04 & 0.88 & 0.16 & 0.37 \\
Colombia & 0 & 0.4 & 0 & 0 & 0 & 0.96 & 0 & 0.19 \\
Peru & 0.16 & 0.16 & 0 & 0.72 & 0 & 0 & 0 & 0.15 \\
Brazil & 0.08 & 0.8 & 0.12 & 0.2 & 0 & 0.2 & 0 & 0.2 \\
France & 0.36 & 0.8 & 0 & 0 & 0.2 & 0.2 & 0.04 & 0.23 \\
Armenia & 0.2 & 0 & 0 & 0 & 0 & 0 & 0 & 0.03 \\
Greece & 0.24 & 0.24 & 0 & 0.48 & 0.2 & 0.2 & 0 & 0.19 \\
Spain & 0.28 & 0 & 0 & 0.12 & 0.44 & 0.88 & 0.48 & 0.31 \\
Germany & 0.08 & 0.8 & 0.2 & 0.6 & 0.4 & 0.2 & 0.16 & 0.35 \\
Portugal & 0.08 & 0.24 & 0 & 0 & 0 & 0.32 & 0 & 0.09 \\
Australia & 0 & 0.24 & 0 & 0.4 & 0.2 & 0.24 & 0 & 0.15 \\
Italy & 0.36 & 0.24 & 0.16 & 0.52 & 0.32 & 0.2 & 0 & 0.26 \\
Russia & 0 & 0.36 & 0.16 & 0.04 & 0.48 & 0.2 & 0 & 0.18 \\
\bottomrule
\end{tabular}%
}
\caption{Fraction of answers in which artifacts of a country are represented, when Gemini 2.5 Pro is asked under specific questions.}
\label{tab:ratio_metric_gemini}

\end{table*}

\begin{table*}[h!]
\centering
\resizebox{\textwidth}{!}{%
\begin{tabular}{l c c c c c c c c}
\toprule
\textbf{Country} & \textbf{Cuisine} & \textbf{Holidays \& Festivals} & \textbf{Clothing \& Accessories} & \textbf{Landmarks} & \textbf{Historical Events} & \textbf{Sportspeople} & \textbf{Sports Teams} & \textbf{Average} \\
\midrule
India & 0.44 & 0.96 & 0.2 & 0.96 & 0.2 & 0 & 0 & 0.39 \\
South Korea & 0.12 & 0.04 & 0.2 & 0 & 0.2 & 0 & 0 & 0.08 \\
China & 0.52 & 0.96 & 0.24 & 1 & 0.24 & 0 & 0 & 0.42 \\
Thailand & 0.2 & 0.72 & 0 & 0 & 0 & 0 & 0 & 0.13 \\
Japan & 0.36 & 0.68 & 0.04 & 0.44 & 0.04 & 0 & 0 & 0.22 \\
Indonesia & 0.04 & 0 & 0.12 & 0.2 & 0.12 & 0 & 0 & 0.07 \\
UAE & 0 & 0.6 & 0.2 & 0.48 & 0.2 & 0 & 0 & 0.21 \\
Iran & 0 & 0.6 & 0 & 0.04 & 0 & 0 & 0 & 0.09 \\
Ethiopia & 0 & 0.6 & 0.12 & 0 & 0.12 & 0 & 0 & 0.12 \\
South Africa & 0.24 & 0.88 & 0.04 & 0.08 & 0.04 & 0 & 0.36 & 0.23 \\
Morocco & 0 & 0.56 & 0.16 & 0 & 0.16 & 0 & 0 & 0.13 \\
Ghana & 0 & 0.64 & 0.16 & 0 & 0.16 & 0 & 0 & 0.14 \\
Nigeria & 0 & 0 & 0.16 & 0 & 0.16 & 0 & 0 & 0.05 \\
Canada & 0.16 & 0.88 & 0 & 0.16 & 0 & 0 & 0.36 & 0.22 \\
Mexico & 0.36 & 0.88 & 0 & 0.28 & 0 & 0 & 0 & 0.22 \\
USA & 0.6 & 0.92 & 0.2 & 1 & 0.2 & 0.24 & 0.52 & 0.53 \\
Argentina & 0.44 & 0.84 & 0.2 & 0.76 & 0.2 & 0.24 & 0 & 0.38 \\
Colombia & 0.24 & 0.76 & 0 & 0 & 0 & 0.24 & 0 & 0.18 \\
Peru & 0.28 & 0.28 & 0 & 1 & 0 & 0 & 0 & 0.22 \\
Brazil & 0.16 & 0.96 & 0 & 0.76 & 0 & 0.04 & 0 & 0.27 \\
France & 0.48 & 0.92 & 0.08 & 0.08 & 0.08 & 0 & 0.08 & 0.25 \\
Armenia & 0.16 & 0.08 & 0 & 0.04 & 0 & 0 & 0 & 0.04 \\
Greece & 0.48 & 0.56 & 0 & 0.88 & 0 & 0 & 0 & 0.27 \\
Spain & 0.44 & 0.08 & 0.16 & 0.72 & 0.16 & 0.24 & 0.4 & 0.31 \\
Germany & 0.08 & 0.96 & 0.2 & 1 & 0.2 & 0 & 0 & 0.35 \\
Portugal & 0.04 & 0.56 & 0.04 & 0 & 0.04 & 0.24 & 0 & 0.13 \\
Australia & 0 & 0.48 & 0.04 & 0.68 & 0.04 & 0 & 0 & 0.18 \\
Italy & 0.2 & 0.56 & 0.16 & 0.96 & 0.16 & 0 & 0.04 & 0.3 \\
Russia & 0.04 & 0.36 & 0.2 & 0.16 & 0.2 & 0 & 0 & 0.14 \\
\bottomrule
\end{tabular}%
}
\caption{Fraction of answers in which artifacts of a country are represented, when GPT-4o is asked under specific questions.}
\label{tab:ratio_metric_gpt}
\end{table*}
\end{document}